\documentclass[11pt,twoside]{article}
\usepackage{asp2004}
\usepackage{psfig}
\usepackage{epsf}
\usepackage{graphics}
\usepackage{lscape}
\def\edcomment#1{\iffalse\marginpar{\raggedright\sl#1\/}\else\relax\fi}
\reversemarginpar

\begin{document}
\title{Conference Summary: Astronomy Perspective of Astro-Statistics}
 \author{Ofer Lahav}
\affil{Department of Physics and Astronomy, 
University College London,
Gower Street, London WC1E 6BT, UK
}

\begin{abstract}
  This is a summary of the `Astronomy Perspective' of the 4th meeting
  on {\it Statistical Challenges in Modern Astronomy} held at Penn
  State University in June 2006.  We comment on trends in the
  Astronomy community towards Bayesian methods and model selection
  criteria.  We describe two examples where Bayesian methods have
  improved our inference: (i) photometric redshift estimation (ii)
  orbital parameters of extra-solar planets.  We also comment on the
  pros and cons of Globalization of scientific research.  Communities
  like Astronomy, High Energy Physics and Statistics develop ideas
  separately, but also have frequent interaction. This illustrates the
  benefits of `comparing notes'.

\end{abstract}

\vspace{-0.5cm}
\section{Introduction}

The dramatic increase of data in Astronomy has renewed  interest in
the principles and applications of statistical inference methods.
These methods can be viewed as a bridge between the data and the
models. For example, the same statistic can be applied to both data and
models as an effective way of comparing the two.  

\noindent
The meeting focused on the following applications in Astronomy:

\begin{itemize} 

\item Cosmology  

\item Small-N problems (including High Energy Physics)

\item Astronomical surveys

\item Planetary systems

\item Periodic variability

%\item Developments in statistics

%\item Cross-disciplinary perspectives

\end{itemize} 

\noindent
The common statistical problems
in these and other  examples from Astronomy and Physics 
fall broadly into the following tasks:

\begin{itemize} 

\item Data compression (e.g. galaxy images or spectra).

\item Classification  (e.g. of stars, galaxies or Gamma Ray Bursts).

\item Reconstruction  (e.g. of blurred galaxy images or mass distribution
      from weak gravitational lensing).

\item Feature extraction 
(e.g. signatures feature of stars, galaxies or quasars). 

\item Parameter estimation (e.g. orbital parameters of extra-solar planets
or cosmological parameters).

\item Model selection (e.g. are there 0,1,2,... planets around a star, 
   or is a cosmological model with non-zero neutrino mass more favourable).

\end{itemize}
 
It is possible for these tasks to be related.  For example, estimation
of cosmological parameters from the Cosmic Microwave Background (CMB) or
galaxy redshift surveys are commonly deduced from a compressed
information, usually in the form of the angular and 3D power spectra,
respectively.  As further example is classification of galaxy
spectra. It can be achieved in a compressed space of the spectra, or
in the space of astrophysical parameters estimated from the spectra.

The Astro-Statistics community is fortunate to have these days
textbooks written by astronomers and physicists, among them (in
chronological order): Lyons (1991), Lupton (1993), Babu \& Feigelson
(1996), Sivia (1996), Cowan (1998), Starck \& Murtagh (2002), Martinez
\& Saar (2002), Press et al. (1992), Wall \& Jenkins (2003), Saha
(2003) and Gregory (2005).
%Press et al.

\section {Inference Methods}

There is an ongoing debate between the `Frequentist' approach
and the `Bayesian' methodology.
The `Frequentist' approach 
interpret  probability as the frequency of the outcome of a
repeatable experiment.
In contrast, the `Bayesian' methodology (first published
in 1764) views the interpretation of probability  more generally and it 
includes a degree of belief, formulated as:
$$ 
P(model | data)=  
P(data | model) P (model) / P(data), 
$$
where on the right hand side the first term is the {\it likelihood},
the second is the {\it prior} and the third is the {\it evidence}.

In the Bayesian approach the choice of {\it priors} may strongly affect 
the inference. However it is an `honest' approach in the sense
that all the assumptions are explicitly spelled out in a logical manner.

\subsection {Sources of Systematics}

A major part of research in Astronomy is devoted to the effect of
systematic errors.  Consider the example of estimating a specific
parameter, e.g. the Dark Energy equation of state parameter $w$ from
Baryon Acoustic Oscillations observed in galaxy clustering (e.g.
Eisenstein et al. 2005).  We can  distinguish three
types of systematics:

\begin{itemize}

\item 
Cosmological uncertainty (due to the assumption on the other 
  $N-1$ cosmological parameters associated priors).

\item
Astrophysical uncertainty (e.g. what is the relation between the clustering
 of luminous galaxies and the matter fluctuations). 

\item
Observational uncertainty (e.g. selection effects in the galaxy sample).

\end{itemize}

Each of these contributes to the error budget of $w$ in a different way 
and should be incorporated in the statistical analysis accordingly.

\subsection {Justifying  Priors}

The choice of prior is crucial in the Bayesian framework, yet the
justification of each prior is not always spelled out in research
articles.  To give an example, a prior on the curvature of the universe
can be justified in a number of ways; some theoretical, some
empirical:

\begin{itemize} 
\item Theoretical  prejudice
  (e.g. `according to Inflation, the universe must be flat').

\item
Previous observations
  (e.g. `we know from the CMB WMAP experiment 
  the universe is flat to  within 2\%, under eh assumion of other priors' ).

\item
 Parameterized ignorance ( e.g. `a uniform prior' 
  or `a Jeffreys prior').

\end{itemize}

\subsection {Recent Trends in Astro-statistics}

We note some recent trends expressed in this, and other conferences.

\begin{itemize} 
 
\item Astro-Statistics has become 
 a `respectable' discipline of its own.

\item
`Bayesian' approaches are more commonly used,
   and in better co-existence with `Frequentist' methods.

\item
There is more awareness of model selection methods, e.g. 
the Akaike Information Criterion (AIC) and the 
Bayesian Information Criterion (BIC), see e.g. Liddle et al. (2006).

\item
Computer intensive methods, e.g. Markov Chain Monte Carlo (MCMC)
   are more popular.

\item
Free software packages are more widely used.

\end{itemize}

It is beyond the scope of this review to cover every topic discussed
at the meeting.  We heard interesting discussions on the role of
statistics in cutting edge research in Astronomy, e.g. the analyses of
the CMB maps (Szapudi, Hinshaw), galaxy
surveys (Martinez, Loredo), micro-lensing (Kochaneck), weak
Gravitational lensing (Bernstein, Lupton), Gravitational waves (Woan),
visualization (Inselberg) and the Virtual Observatories (Hanisch).  In
this short summary I shall focus on two examples; photometric
redshifts and extrasolar planets. Both cases illustrate how Bayesian
approaches have improved our inference on the science questions of
interest.

\section {An Example From Cosmology: Photometric Redshifts}

Mapping the galaxy distribution in 3D requires the galaxy redshifts.
In the absence of spectroscopic data, redshifts of galaxies may be
estimated using multi-band photometry, which may be thought of as very
low-resolution spectroscopy.  While the redshift error per galaxy is
relatively large, having a great number of galaxies could reduce the
errors on measures of the galaxy clustering.  Photometric surveys over
large areas of the sky may compete well with spectroscopic surveys.
Photo-z methods proved very useful e.g. in  recent analyses of the
GOODS, COMBO-17 and SDSS Luminous Red Galaxies.  
%For recent analyses of the LRG photometric
%redshift surveys see e.g.  Blake et al. (2006) and Padmanabhan et
%al. (2006).  
Several wide-field photometric redshifts are planned for
the future, e.g. the Dark Energy Survey, PanSTARRS and LSST.
Understanding the photometric redshift errors is crucial for quantifying the 
errors on e.g. the Dark Energy equation of state parameter $w$
from galaxy clustering or weak lensing. 

In more detail, photometric redshift methods rely on measuring the
signal in the photometric data arising from prominent "break" features
present in galaxy spectra e.g. the 4000 $\AA$ break in red, early-type
galaxies, or the Lyman break at 912 $\AA$ in blue, star-forming
galaxies.  There are two basic approaches to measuring a galaxy
photometric redshift $z$ (e.g. Csabai et al. 2003 and references
therein).  The first, template matching, relies on fitting model
galaxy spectral energy distributions (SEDs) to the photometric data,
where the models span a range of expected galaxy redshifts and
spectral types.  This is done via a simple $\chi^2$ statistic, i.e.
via the likelihood $P(colours|z)$, but it may lead to catastrophic
errors.  Benitez (2000) generalized the method by incorporating
Bayesian priors.  The prior $P(z|magnitude)$ for the redshift of a
galaxy given its magnitude (apparent luminosity) then multiplies the
likelihood to give the posterior
$$
P(z|colours, magnitude) \propto 
P(colours|z) \times P(z|magnitude) \; .
$$
This Bayesian chain, which can also be generalized to include galaxy type, 
greatly reduces the number of outliers.

Another approach utlilises an existing spectroscopic redshift sample
as a training set to derive an empirical photometric redshift fitting
relation.  An example of a training-based method, ANNz, which is also
Bayesian, utilizes Artificial Neural Networks (Collister \& Lahav
2004).
%and is 
%freely available\footnote{http://www.star.ucl.ac.uk/$\sim$lahav/annz.html}.
When applied to SDSS galaxies the rms error using ANNz is $\sigma_z =
0.02$, compared with $\sigma_z = 0.07$ using a template method.

\section{An Example From Extra-solar Planets: Orbital Parameter Estimation}

Astronomers have faced a growing number of free parameters in
modelling astrophysical systems, for example  cosmological parameters or
extra-solar planet orbital parameters. In the case of a model with $N$
free parameters marginalizing over $N$-1 parameters, it proves to be
 computationally
expensive if the parameter space is mapped into a grid.  An
alternative method, the Markov Chain Monte Carlo (MCMC), has been
known since the 1950's and a wide range of methods exists in the
literature to implement it, e.g. the Metropolis-Hasting algorithm.

The key idea is to turn a probability distribution function in $N$
dimensions into a cloud of points which represents the probability
distribution function.  The probability distribution function could
incorporate the probabilities for the priors, in the Bayesian spirit.
The MCMC algorithm constructs a random walk in the model parameter
space with steps drawn from a multi-dimensional proposal
distribution (e.g. a Gaussian).  It is crucial to apply tests for
convergence, i.e. to ensure that the parameter space is properly
sampled, in particular if there are several peaks in a highly
dimensional space.
 
MCMC algorithms have been applied recently to parameter estimation
from the CMB and other cosmological data sets
(e.g. Lewis \& Bridle 2003; Verde et al. 2003) and to both detecting
and characterizing orbits of extrasolar planets (e.g. Gregory 
2005; Ford 2005; and Ford and Gregory in this volume).

Nearly 200 extrasolar planets have been discovered over the past
decade.  Most of those were discovered using measurements of the
radial velocity of the host star.  The radial velocity curve can be
modelled by approximately a dozen parameters, depending on the
complexity of the assumed model.  It is also important to allow for
more than one planet around the star, hence for more free parameters.
This leads to the challenging problem of non-linear minimization in a
highly dimensional parameter space.  Deriving these parameters
accurately is very important as this can then influence the
interpretation for an individual object, as well as the statistics of
orbital parameters for an ensemble of extra-solar planets.

For example, in many of the discovery papers the approach taken is to
estimate first the period $P$ and then for that fixed $P$ to solve
later for the orbital parameters.  As there is degeneracy of parameters
and dependence on their priors this could lead to the wrong value of
$P$. This was pointed out by Gregory (2005), who developed an MCMC
Bayesian approach to cope with the
multi-parameter estimation.  He illustrated the method for the data
for HD73526, where he found three possible solutions for P. In fact
the previously reported one turned out to be the least probable orbit
(but apparently the data for this system somewhat changed since the publication
of the paper).

\section {Future Work in Astro-statistics} 

Based on the discussions during the meeting,
the following topics call attention 
call for further work in Astro-statistics:

\begin{itemize} 

\item
Model selection methodology (e.g. which criteria and the role of 
priors).

\item
MCMC machinery and extensions (e.g. nested sampling).

\item
Detection of non-Gaussianity and shape finders (e.g. for galaxy survey and
CMB maps).

\item
Blind de-convolution (e.g. for recovering galaxy shapes from blurred images).

\item
Object classification (e.g. stars, galaxies and quasars).

\item
Comparing simulations with data (e.g. large galaxy surveys with N-body 
and hydrodynamic simulations).

\item
Visualization (of e.g. 3D galaxy surveys or multi-parameter space). 

\item Virtual Observatories (including both  Real Data and Mock data).

\end{itemize}

\section {The Globalization of Scientific Research}

We shall discuss now the pros and cons of information exchange between 
scientific communities, first communities with common scientific interests 
which were separated by political 
boundaries, and then communities which have different research interests.

The society at large is going through a `Globalization' process.
There is a diversity of definitions for Globalization, some in
positive context, others with negative connotations.  The sociologist
Anthony Giddens defines Globalization as 
 ``decoupling of space and
time - emphasizing that with instantaneous communications, knowledge
and culture can be shared around the world simultaneously\footnote{http://globalisationguide.org/}.''  
Another
definition given in the same website sees Globalization  as being 
``an
undeniably capitalist process. It has taken off as a concept in the
wake of the collapse of the Soviet Union and of socialism as a viable
alternate form of economic organization.''  

A further discussion on Globalization can be seen through  
Thomas Friedman's book
(2005) ``the World is Flat'' (indeed an interesting title in the
context of Cosmology!).
He questions whether ``the world has got too small
and too flat for us to adjust''.

Research in academia is of course a human activity affected, like any
other sector, by the social and technological changes and trends.  The
advantages of Globalization for academic research are numerous: open
access to data sources for all (e.g. via the World Wide Web), rapid  exchange
of ideas, and international research teams.  These aspects make
science more democratic and they enable faster achievements in
`searching for the truth'.  The numerous conferences, electronic
archives and tele-conferences generate a `global village of thinkers'.
While this could lead to a faster convergence in answering fundamental
questions, there is also the risk of preventing independent and original
ideas from developing, as most researchers might be {\it too influenced by
the consensus view}.

Let us consider the current popular cosmological `concordance' model.
Cosmological measurements of Supernovae Type Ia, Large Scale Structure
and the CMB WMAP3 (e.g. Spergel et al. 2006)
are consistent with a `concordance' model in which the universe is
flat and contains approximately 4\% baryons, 21\% dark matter and 75\%
dark energy.  However the two main ingredients, Dark Matter and Dark
Energy, are still poorly understood. We do not know if they are `real'
or they are the modern `epicycles', i.e. `fudge factors' which help to
fit the data better, until a new theory will greatly simplify our
understanding of the observations.

A disturbing question is whether the popular cosmological `concordance
model' is a result of Globalization?  It is interesting to contrast the
present day research in Cosmology with the research in the 1970's and
1980's.  This was the period of the `Cold War' between the former
Soviet Union and the West.  During the 1970's the Russian school of
Cosmology, led by Y. Zeldovich, advocated massive neutrinos, `Hot Dark
Matter', as the prime candidate for dark matter.  As neutrinos were
relativistic when they decoupled, they wiped out structure of small
scales.  This led to the `top-down' scenario of structure formation.
In this picture `Zeldovich pancakes' of the size of superclusters
formed first, and then they fragmented into clusters and galaxies.
This was in conflict with observations, and cosmologists concluded
that neutrinos are not massive enough to make up all of the dark
matter.  The downfall of the top-down `Hot Dark Matter' scenario of
structure formation, and the lack of evidence for neutrino masses from
terrestrial experiments made this model unpopular.  The Western school
of Cosmology, led by J. Peebles and others, advocated a 'bottom up'
scenario, in framework which later became known as the popular Cold
Dark Matter.  However the detection of neutrino oscillations showed
that neutrinos indeed have a mass, i.e. Hot Dark Matter does exist,
even if in small quantities. Current upper limits are in the range
between 4\% (from LSS or CMB alone) to 0.5\% (from a combination of
cosmic probes).  therefore both forms, Cold Dark Matter and Hot Dark Matter,
may exist in nature. This example illustrates that having two independent
schools of thoughts was actually beneficial for progress in Cosmology. 

Luckily, despite the current global village effect, we still have communities
that work independently, not geographically or politically, but in
different areas of research.  Astronomy, High Energy Physics and
Statistics are examples of such independent communities.  Fortunately,
these three communities talk to each other from time to time to
`compare notes'.  This meeting provided an excellent example of such
fruitful interaction.  Fundamental issues in statistical inference
from data will not go away.  With the exponential growth of data in
Astronomy there is a great need for further interaction of astronomers with
experts in other fields.

\begin{quote}
{\bfseries Acknowledgements.} 
It is a pleasure to thank, on behalf of the participants, the
Co-organizers Jogesh Babu, Eric Feigelson, the SOC (JB, EF, Jim
Berger, Kris Gorski, Thomas Laredo, Vicent Martinez, Larry Wasserman,
Michael Woodroofe), the Graduate students (Hyunsook Lee, Derek Young),
the Conference Planner (John Farris) and the Sponsors (SAMSI, NSF,
NASA, IMS, PSU). I also thank Filipe Abdalla, Sarah Bridle, Anais Rassat and
Kate Wilson-Heyworth for their comments on this manuscript.

\end{quote}

%OL's definitions

\def \aap{A\&A}
\def \apj{ApJ}
\def \mnras{MNRAS}
\def \pasa{Publ. Astron. Soc. Austral.}
\def \prep{in preparation}
\def \prl{Phys. Rev. Lett.}
\def \pra{Phys. Rev. A}
\def \aj{AJ}


\begin{thebibliography}{}
%
%\addcontentsline{toc}{section}{References}

\bibitem{BF96} Babu, G.J., Feigelson, E.D., 1996,
{\it Astrostatistics},
Chapman \& Hall/CRC

\bibitem{benitz} Benitz, N., 2000, ApJ, 536, 571

\bibitem{cl04} Collister, A. \&  Lahav, O., 2004,  PASP 116, 345 


\bibitem{Cowan} 
Cowan, G.,1998 
 {\it Statistical Data Analysis}, 
Oxford University Press

\bibitem{csabai03} Csabai, I., et al. 2003, AJ, 125, 580  

\bibitem{E05}
Eisenstein, D.J. et al., 2005, ApJ, 633, 560
\bibitem{ford05} Ford, E.B., 2005, astro-ph/0512634  

\bibitem{F05} Friedman, T.L., 2005, {\it The World is Flat}, 
Penguin, Allen Lane.

\bibitem{Gregory05} Gregory, P.C., 2005, 
{\it Bayesian Logical Data Analysis for the Physical Sciences}, 
Cambridge University Press

  
%\bibitem{lahav} Lahav, O., 2001, in Mining the Sky, Proceedings of the
%  MPA/ESO/MPE Workshop held at Garching, Germany, Edited by A. J.
%  Banday, S. Zaroubi, and M. Bartelmann. Heidelberg: Springer-Verlag,
%  2001., p.33 (astro-ph/0012407)

\bibitem{lb02} Lewis, A. \& Bridle, S.L., 2002, PRD, 66, 103511 

\bibitem{liddle06} Liddle, A.R., Mukherjee, P. \& Parkinson, D., 
astro-ph/0608184

\bibitem{Lupton} Lupton, R., 1993,
{\it Statistics in Theory and in Practice}, 
Princeton University Press

\bibitem{Lyons} Lyons, L., 1991,
{\it A Practical Guide to Data Analysis for Physical 
Sciences Students}, 
Cambridge University Press


%\bibitem{Murtagh87}{Murtagh} F., {Heck} A., 1987.\newblock {\it
%    Multivariate data analysis}, Astrophysics and Space Science
%  Library, Reidel, Dordrecht.


\bibitem{MS02}
Martinez, V.J., Saar, E., 2002,
{\it Statistics of the Galaxy Distribution}, 
Chapman \& Hall/CRC

\bibitem{P92}
Press, W.H. et al., 1992, 
{\it Numerical Recipes}, Cambridge University Press

\bibitem{Saha}
Saha, P., {\it Principles of Data Analysis}, 2006
Available free on the WWW.

\bibitem{Sivia96} Sivia, D., 1996,
{\it Data Analysis: A Bayesian Tutorial}, 
Oxford University Press

\bibitem{WMAP3}
Spergel, D.N et al., 2006, ApJ, submitted, astro-ph/0603449 

\bibitem{Starck02}
Starck, J-L, Murtagh, F., 2002, 
{\it Astronomical Image and Data Analysis}, 

  
\bibitem{Verde} Verde, L. \& the WMAP Team, 2003, ApJS, 148, 196

\bibitem{Wall03} Wall, J.V, Jenkins, C.R., 2003,  
{\it Practical Statistics for Astronomers},   
Cambridge University Press


  
\end{thebibliography}
\end{document}